\documentclass[prints]{aa}
\usepackage{psfig}
\usepackage{supertabular}
\usepackage{graphicx}
\begin{document}

\title{H$\alpha$ imaging of the Local Volume galaxies.
		       I. The NGC 6946 galaxy group.}
\titlerunning{H$\alpha$ imaging of galaxies arround NGC 6946}

\author{I.D.Karachentsev\inst{1}
\and S.S.Kajsin\inst{1}
\and Z.Tsvetanov\inst{2}
\and H.Ford\inst{2}}

\institute{Special Astrophysical Observatory, Russian Academy of Sciences,
	 N.Arkhyz, KChR, 369167, Russia
\and Department of Physics and Astronomy, Johns Hopkins University,
     Baltimore, MD 21218, USA}
\date{}

\abstract{ We present new H$\alpha$ imaging of all known dwarf
  irregular companions to NGC~6946: UGC~11583, KK~251, KK~252, KKR~55,
  KKR~56, Cepheus~1, KKR~59, and KKR~60. The galaxies span a range of
  blue absolute magnitudes of [$-$13.6,~$-$17.6], relative gas content
  of [0.1, 2.5] $M_{\sun}/L_{\sun}$, current star formation activity
  of [0.2, 5.2]$10^{-2} M_{\sun}$ yr$^{-1}$, and timescale to exhaust the
  current gas supply of [6, 86] Gyr.
}
\maketitle

\section{ Introduction}

In a volume-limited sample, dwarf galaxies constitute the vast
majority of the total galaxy population. Yet, their star formation
histories remain poorly understood. Up until 2003, among the 362
known neighboring galaxies with distances $D <$ 8 Mpc, of which dwarfs
constitute about 85\% (Karachentsev et al. 2004), only $\sim$15\% have
been imaged in the fundamental H$\alpha$ line. Recent observations
obtained by Gil de Paz et al. (2003), James et al.(2004), Helmboldt et
al.  (2004), and Hunter \& Elmegreen (2004) significantly advanced our
understanding of star formation properties of nearby dwarf galaxies.
Nevertheless, the present knowledge of the local star formation rate,
determined by a cumulative H$\alpha$ flux of the galaxies in the Local
Volume (LV), remains very incomplete.  For a substantial improvement
of this situation, we started a program to obtain H$\alpha$ images of
all 214 northern (DEC$ > 0^{\circ}$) dwarf galaxies within 8 Mpc,
having angular diameters $a_{25} < 8\arcmin$ compatible with the SAO
6-m telescope CCD camera field of view. We believe that this project
will also be supported by systematic H$\alpha$ imaging of all southern
neighboring dwarf galaxies. As a result, we anticipate to get the
first complete data on the H$\alpha$ luminosity function of galaxies
in the Local Universe and the total H$\alpha$ flux of all galaxies
within a distance of 8 Mpc.

In addition, we intend to prepare the first Atlas of H$\alpha$ images
for a complete sample of the LV galaxies, which can be compared with
existing sets of their HI maps. The creation of such a complete
reference sample at zero redshift is fundamental for studying the
evolution of star formation in distant galaxy samples (like the Hubble
Deep Field), as well as the relation between star formation activity
and galaxy environment. We note that many LV galaxies are objects that
have been discovered only recently. In particular, about a hundred
such galaxies were found on POSS-II plates by Karachentseva and her
co-workers over the last few years.  Most of the LV galaxies are
targets of the ongoing Snapshot Survey with the Wide Field and
Planetary Camera 2 and Advanced Camera for Surveys aboard the Hubble
Space Telescope (programs 8192, 8601, 9771 and 10235), for which
accurate distances are determined using the tip of the red giants
branch method. This gives us a unique possibility to compare their
H$\alpha$ structure with high resolution images in the $V,I$ bands and
correctly determine their luminosities.

Here, we present the first results from our H$\alpha$ imaging
program for all 8 dwarf irregular (dIr) companions to the bright
spiral galaxy NGC~6946. In follow up papers, we will study the rate of
star formation in other nearest groups, as well as in the general
local field.

\section{ Observations and data reduction}

We obtained H$\alpha$ and red continuum CCD images for the eight
known members of the NGC 6946 group during 3 observing runs between
September 2001 and November 2002. A typical seeing was about 1.7
$\arcsec$. These observations were carried out at the
SAO 6-meter telescope using an imaging camera equipped with a
2048$\times$2048 pixel CCD. The pixel scale was 0.25$\arcsec$/pixel.
The area imaged was $6\arcmin\times6\arcmin$. The H$\alpha$ + [NII]
emission line fluxes were obtained by observing each galaxy through
two filters: a narrow-band ($\sim$70 \AA) interference filter centered
on 6567 \AA, and a broad $R$-band filter (SED707, $\lambda_{cen}$ = 6780
\AA, $\Delta \lambda$ = 1100 \AA) to determine the nearby continuum
level. Integration times were typically 2$\times$300 sec in the $R$
band and 2$\times$600 sec in H$\alpha$.
Because of a small range of radial velocities, we use the
same H$\alpha$ filter for all objects.

The data reduction was performed using standard MIDAS package tasks.
The following steps were taken to obtain continuum-free H$\alpha$ +
[NII] images.  First, the images were bias subtracted and flat fielded
by a median twilight sky flat taken through the appropriate
filter. After automatic cosmic rays removal, the sky subtraction
was done by fitting a surface underneath the object of
interest using the areas away from the object. Because the seeing
was slightly different in different images, we match the seeing by
degrading all the images to the worst one.

Finally, to obtain  H$\alpha$+[NII] continuum-free images, the
$R$-band images were scaled relative to the interference filter images
using 10--15 unsaturated stars, and then subtracted from the
interference filter images.  Flux calibration for the images was made
with the observations of Oke (1990) standard stars.
The typical errors in total H$\alpha$ fluxes
due to background sky subtraction and photometric calibration are
within 10\% (apart from the KK 251 observed during a non-photometric
night).

Figure 1 presents (from left to right) the H$\alpha$ on-line,
and the continuum-free images of the eight observed galaxies.
Residuals in the continuum-subtracted images due to the
presence of very bright saturated field stars were removed
interactively.

\renewcommand{\tabcolsep}{2pt}
\begin{table*}[t]
\caption{Basic properties of the galaxies. }
\begin{tabular}{p{0.6in}ccrrcccccccl} \hline
Object & $a$  &$b/a$ & $B_t$   &$V_h$   &$V_{LG}$ & $M_B$  &$M_{HI}/L$  &$A(H\alpha)$&  $F_{H\alpha}\times10^{13}$     &$F_c\times10^{13}$   &SFR   &$t_{gas}$ \\
\hline
       &$\arcmin$&    &   mag& km s$^{-1}$&km s$^{-1}$&  mag & $M_{\sun}/L_{\sun}$ &  mag &  erg s$^{-1}$ cm$^{-2}$ &  erg s$^{-1}$ cm$^{-2}$  &$M_{\sun}$ yr$^{-1}$ &  Gyr  \\
\hline
(1)   &  (2)  &  (3)  &  (4)  &  (5)  &  (6)  &  (7)  &  (8)  &  (9)  &  (10)  &  (11)  &  (12)  &(13) \\
\hline
UGC~11583 & 1.8& .44& 15.90& 127 &430 &$-$14.28&  2.04 &  0.71&  0.29$\pm$.04&  0.56& 0.0025&  86   \\
       &    &    &      &     &    &      &       &      &           &      &      &        \\
KK~251  & 1.6& .50& 16.5 & 130 &433 &$-$13.63&  2.41 &  0.69&  0.31$\pm$.20&  0.58& 0.0026&  54   \\
       &    &    &      &     &    &      &       &      &           &      &      &        \\
KK~252  & 0.9& .99& 16.7 & 138 &441 &$-$14.11&  0.10 &  1.05&  0.16$\pm$.03&  0.42& 0.0018&  8.4  \\
       &    &    &      &     &    &      &       &      &           &      &      &        \\
NGC~6946  &11.5& .85&  9.58&  51 &355 &$-$20.86&  0.21 &  0.80& 339      & 708 & 3.12 &  3.1   \\
       &    &    &      &     &    &      &       &      &           &      &      &        \\
KKR~55  & 0.6& .67& 17.0 &  32 &337 &$-$14.79&  0.34 &  1.58&  0.48$\pm$.04&  2.06& 0.0091&  6.3  \\
       &    &    &      &     &    &      &       &      &           &      &      &        \\
KKR~56  & 0.7& .64& 17.6 & -47 &260 &$-$14.39&  0.52 &  1.69&  0.10$\pm$.09&  0.47& 0.0021&  29   \\
       &    &    &      &     &    &      &       &      &           &      &      &        \\
Ceph~1  & 3.0& .50& 15.4 &  58 &367 &$-$17.53&  0.70 &  2.18&  0.66$\pm$.18&  4.92& 0.0216&  67   \\
       &    &    &      &     &    &      &       &      &           &      &      &        \\
KKR~59  & 2.3& .61& 15.7 &   1 &311 &$-$17.01&  0.30 &  2.08&  1.75$\pm$.23& 11.9 & 0.0523&  7.4  \\
       &    &    &      &     &    &      &       &      &           &      &      &        \\
KKR~60  & 0.7& .70& 18.  & -35 &275 &$-$15.42&   ---   &  2.46&  0.84$\pm$.27&  8.10& 0.0356&   ---    \\
\hline
\end{tabular}
\end{table*}

\section{Results}

The basic parameters of nine galaxies in the NGC~6946 group are listed
in Table. Column~1 contains the number of each galaxy in the Uppsala
Galaxy Catalogue (UGC) or in the lists by Karachentseva \&
Karachentsev (1998, KK) and Karachentseva et al. (1999, KKR); columns
2,3 gives major angular diameter in arcminutes and apparent axial
ratio measured at a level of $\sim25^m/\sq\arcsec$; column 4 presents
the integrated apparent $B$-magnitude; columns 5,6 give the
heliocentric radial velocity derived by Makarov et al. (2003) or
Huchtmeier et al.  (2000a,b, 2003), and the radial velocity with
respect to the Local Group centroid under the apex parameters
according to Karachentsev \& Makarov (1996); column 7 presents the
absolute magnitude corrected for the Galactic extinction, $A_B$,
according to Schlegel et al. (1998); column 8 lists the hydrogen
mass-to-blue luminosity ratio from Karachentsev et al.(2004); column 9
presents the Galactic extinction $A(H\alpha) = 0.538 A_B$; columns
10,11 give the observed total H$\alpha$ + [NII] flux with
its standard error and the total flux corrected for the Galactic
extinction; column 12 lists the star formation rate, SFR($M_{\sun}$
yr$^{-1}$) = 1.27$\times10^9 F(H\alpha) D^2$, where $D$ is the galaxy
distance in Mpc, and $F(H\alpha)$ is in units of erg
s$^{-1}$ cm$^{-2}$; and the last column gives the
time (in Gyr) to exhaust the total current gas content of
the galaxy at the current star formation rate, (Total Gas Mass)/(SFR),
where the total gas mass is 1.32 $\times$ M(HI) to account for He. For
the bright spiral NGC~6946 its H$\alpha$ flux in column 10 was taken
from Young et al. (1996), corresponding to the inner 6 arcmin region
only.

Almost all the group galaxies have been resolved into stars by Sharina
et al. (1997) and Karachentsev et al. (2000), who determined the mean
distance to the group via the brightest stars to be (5.9$\pm$0.4)
Mpc. This value was used by us to calculate the galaxy absolute
magnitudes and SFRs given in Table.

\section{ Discussion}

Until the 1990s, the giant spiral NGC~6946 was considered as an
isolated galaxy situated on the edge of the Local Void (Tully,
1988). Recent discoveries of eight companions establish NGC~6946 as
the brightest member of a loose group. The root mean relative velocity
of companions is ($-2\pm24$) km s$^{-1}$ , showing no asymmetry with
respect to NGC~6946. The mean square radial velocity dispersion of the
companions, 64 km s$^{-1}$, and the mean projected linear separations,
245 kpc, are typical for a loose group dominated by a late type
galaxy. The virial mass-to-blue luminosity ratio for the NGC~6946
group, 30 $M_{\sun}/L_{\sun}$, is also characteristic for the nearby
groups.

Ferguson et al. (1998) presented deep H$\alpha$ image of NGC~6946, which
reveal the presence of HII regions out to two standard optical radii,
$R_{25}$.  Young et al. (1996) determined the total H$\alpha$ flux
from the inner (6$\arcmin$) region of the galaxy. Comparing the data
on column 11, we find that 96\% of the total H$\alpha$ flux
of the NGC~6946 group comes from its brightest spiral galaxy.  This
result is consistent with the conclusion of Nakamura et al.  (2003)
based on the Sloan Digital Sky Survey that 83\% of the blue luminosity
density comes from spiral galaxies, 5\% from irregular galaxies and
9\% from early type ones. Therefore, in the present epoch, the disks
of massive spiral galaxies are the basic hearths of current star
formation activity.

The distribution of the NGC 6946 companions in integrated star
formation rate spans a range from 0.002 to 0.052 in $M_{\sun}$
yr$^{-1}$ that is typical for irregular galaxies (see Fig. 5
in Hunter \& Elmegreen, 2004).  The median timescale to exhaust the
current gas supply for the dIr companions to NGC~6946, $t_{gas}$ = 29
Gyr, is one order longer than the timescale for the NGC~6946 itself. The
gas depletion timescale $t_{gas}$ for the group members is also typical
for the sample of dIr galaxies in (Hunter \& Elmegreen, 2004).

\acknowledgement{ This research was supported by RFFI--DFG grant
  02--02--04012 and RFFI grant 04--02--16115.}

{}

\begin{figure*}
\includegraphics[width=14cm]{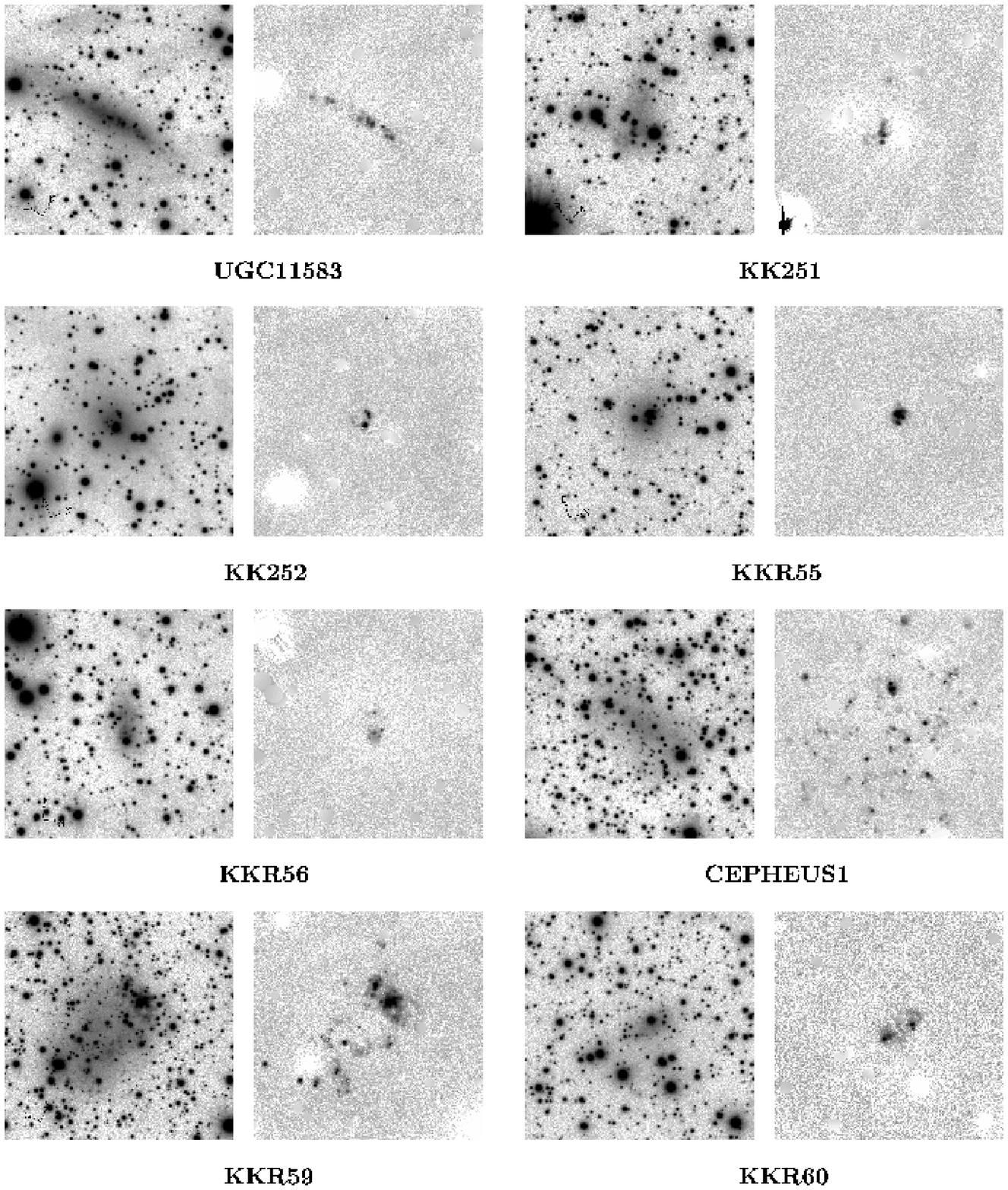}
\caption{From left to right: the H$\alpha$,  and the continuum-free
images of eight observed galaxies around NGC~6946.
{\em \bf UGC~11583.} The large-scale image of this elongated
irregular galaxy (Karachentsev et al. 2000) reveals
several stellar associations in its body. The continuum-subtracted
H$\alpha$ image (top right in Fig. 1) shows a dozen of compact
HII-regions aligned along the major axis. We did not find any
appreciable diffuse H$\alpha$ emission in UGC~11583 with an upper
limit of about 0.05 10$^{-13}$ erg s$^{-1}$ cm$^{-2}$.
{\em \bf KK~251.} This dIr galaxy is situated only 6$\arcmin$ away from
UGC~11583.  Its continuum-subtracted H$\alpha$ image shows three
compact HII-knots and several diffuse emission regions on the galaxy
southern side.
{\em \bf KK~252.} The well-resolved face-on dIrr galaxy (Karachentsev et
al. 2000). Four very compact HII-regions are seen in its central
part as well as faint diffuse emission at the southern side.
{\em \bf KKR~55.} It may be considered as a blue compact dwarf (BCD)
galaxy with a relatively bright stellar association at $8\arcsec$ NE
off the center.  The continuum-subtracted H$\alpha$ image of KKR~55
finds a strong knotted HII-region, which coincides with the
association. No appreciable diffuse emission is seen in the remaining
parts of the galaxy.
{\em \bf KKR~56.} The well-resolved dIrr galaxy (Karachentsev et al. 2000).
The continuum-subtracted H$\alpha$ image shows several almost
stellar-like faint knots embedded into a faint diffuse envelope.
{\em \bf Cepheus~1.} This Sm-type very obscured galaxy has been discovered
in HI by Burton et al. (1999), who estimated the galaxy distance as
6.0 Mpc based on the luminosity of its HII regions. The
continuum-subtracted H$\alpha$ image obtained by us confirms the
presence of many isolated compact and diffuse HII regions scattered
over the area much exceeding the visible galaxy body. Probably, the
central South-to-North elongated part of Cepheus1 seen in continuum is
only a brighter bar-like structure of a low surface brightess disk
outlined by HII regions.
{\em \bf KKR~59.} The resolved irregular galaxy has two large ``curly''
HII regions on its northern edge. Several other diffuse emission
knots are seen over the galaxy body. This is the most H$\alpha$
luminous dwarf companion to NGC~6946.
{\em \bf KKR~60.} Unlike other galaxies considered here, KKR~60 is not
resolved into stars so far. The continuum-subtracted H$\alpha$ image
shows several compact HII regions within a diffuse emission
component. The heliocentric H$\alpha$ radial velocity
of KKR~60,  $-$35 km s$^{-1}$, is typical one for the group members.}
\end{figure*}
\end{document}